\documentclass[twocolumn,showpacs,preprintnumbers,amsmath,amssymb,floatfix]{revtex4}

\usepackage{graphicx} 
\usepackage{bm}       

\newcommand{\e}{{\rm e}}
\renewcommand{\i}{i}
\newcommand{\ignore}[1]{}
\providecommand{\Ev}[1]{\langle #1 \rangle}

\providecommand{\Var}[1]{{\rm var}\!\left [#1 \right]}

\providecommand{\myeq}[1]{Eq.~\eqref{#1}}
\providecommand{\myfig}[1]{Fig.~\ref{#1}}
\providecommand{\avg}[1]{{#1}_{\rm avg}}
\providecommand{\given}{\,|}
\providecommand{\Size}{{S}}
\providecommand{\Number}{{N}}

\begin{document}

\title{A transactional theory of fluctuations in company size}
\author{Aaron O. Schweiger}
\email{aschweig@physics.bu.edu}
\affiliation{Center for Computational Science
             and Department of Physics, Boston University,
             Boston, Massachusetts 02215}
\author{Sergey V. Buldyrev}
\affiliation{
            Department of Physics, Yeshiva University,
            New York, New York 10033}
\author{H. Eugene Stanley}
\affiliation{Center for Polymer Studies
             and Department of Physics, Boston University,
             Boston, Massachusetts 02215}


\begin{abstract}

    Detailed empirical studies of publicly traded business firms have
    established that the standard deviation of annual sales growth rates decreases
    with increasing firm sales as a power law, and that the sales growth distribution
    is non-Gaussian with slowly decaying tails.  To explain these empirical facts,
    a theory is developed that incorporates both the fluctuations of a
    single firm's sales and the statistical differences among many firms.
    The theory reproduces both the scaling in the standard deviation
    and the non-Gaussian distribution of growth rates.  Earlier models reproduce the same
    empirical features by splitting firms into somewhat ambiguous
    subunits; by decomposing total sales into individual transactions,
    this ambiguity is removed.  The theory yields verifiable predictions and
    accommodates any form of business organization within a firm.
    Furthermore, because transactions are fundamental to economic activity at all scales,
    the theory can be extended to all levels of the economy,
    from individual products to multinational corporations.


\end{abstract}

\pacs{89.65.Gh,89.75.Da,87.23.-n}

\maketitle \date{\today}
    \section{Introduction}
%
%
%
%
%
	In 1931, Gibrat introduced a formal model of market structure based
    on his {\it Law of Proportionate Effect}, which describes the time evolution of
    business firm size~\cite{Gibrat, Sutton1}.
    There is no unique measure of the size of a company;
    a firm's size is frequently defined to be the total annual sales.
    Gibrat postulated that all firms, regardless of size, have an equal probability to grow a
    fixed fraction in a given period, i.e., the firm size
    undergoes a multiplicative random walk.
    Over appropriately chosen
    periods, the distribution of the annual fractional size change
    is approximately Gaussian, independent of the firm size~\cite{Sutton1}.
    Gibrat hypothesized that the distribution of firm sizes
    follows a nonstationary lognormal distribution.
    Subsequent statistical studies~\cite{Hart,Stanley1} have found that
    the firm-size distribution is well-approximated by a lognormal
    distribution as predicted.
    More recent studies~\cite{Nature96,Scaling1,Plerou2,BottazziPammolli,Fabritiis,Gaffeo}
    have also demonstrated three empirical facts incompatible with Gibrat's hypothesis:
%
    (i)
        Firm growth rates, defined to be the annual change in
        the logarithm of sales, follow a non-Gaussian distribution with
        slowly decaying tails,
    (ii)
        The standard deviation of growth rates scales as a power law with firm
        sales, e.g., ${\sigma_g \sim \Size^{-\beta}}$,
        where ${\beta \approx 0.2}$~\cite{Nature96},
    (iii)
        The firm size distribution is approximately stationary over tens of years.

\begin{figure}
    \includegraphics[height=2.0in,width=3in]{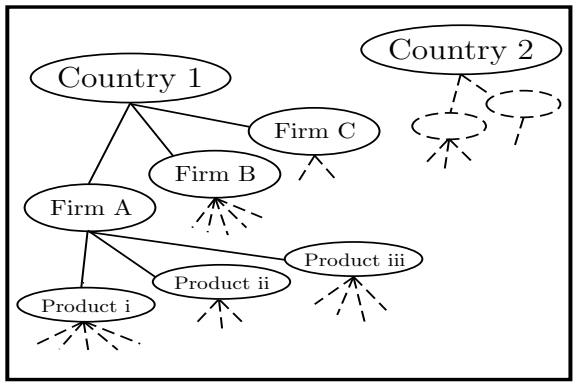}
    \caption{ \label{TheLevels}
    The GDP of a country is decomposed into the contributions
    of individual firms.  Firms' contributions
    are further split by individual products, {\it ad infinitum}.
    Because complex behavior is observed at all scales, models of economies
    that explain complex behavior in terms of fluctuating subunits must account for
    the composite nature of the subunits themselves. }
    \end{figure}

    A number of models have been proposed to explain
    these incompatibilities between Gibrat's theory and the empirical facts.
    These models decompose each firm into fluctuating subunits~\cite{Scaling2,PRL98,Sutton,BottazziPhysica,Wyart,Matia2,Fu,BottazziSecchi,BottazziSecchi2},
    which are usually taken to be management divisions or product lines.
    While the subunits obey simple dynamics, the aggregate
    entities approximate the empirically observed statistics.
%
    Studies ranging from the gross domestic product (GDP) of countries to the
    sales of individual products have found comparable scaling laws and non-Gaussian
    distributions~\cite{Canning98,Lee98,BottazziSecchi,BottazziSecchi2,Matia2,Fu}.
    While the statistics of GDP growth may be explained in terms of the contributions
    of individual firms, the firm sales do not obey simple dynamics.
    Likewise, while firm sales growth may be explained in terms of the
    contributions from products, products exhibit complex
    behavior similar to the entire firm~\cite{Matia2}.
    Empirically, no economic entity obeying simple dynamics
    has been identified.
    In order to reconcile the empirical results of Refs.~\cite{Canning98,Lee98,BottazziSecchi,BottazziSecchi2,Matia2,Fu}
    with these theoretical
    models, either a subunit obeying a postulated simple dynamics must be identified,
    or we must account for the complex nature of the subunits
    themselves, as illustrated by the hierarchal schematic in \myfig{TheLevels}.
%
%
%

    Here we develop a theory to explain the observed firm growth
    statistics that does not invoke an inherently hierarchical model
    of the economy~\cite{Nature96,Scaling2,Buldyrev}.
    We do not make any assumptions about the internal structure of firms.
    We postulate that the total annual sales of a particular firm can be broken
    down into a finite sum of individual sales transactions.
%
%
%
%
    In our analysis, each firm's total annual sales is an independent
    random variable that is characterized by three parameters (see below).
    Furthermore, these parameters vary randomly from firm to firm.
    We refer to the set of firms as the population. We define heterogeneity as
    a measure of the variability of the firms' parameters within a population.

    In Section~\ref{StatisticsOfAFirm}, we develop a model of
    a single firm's sales.  In Section~\ref{FirmHeterogeneity},
    we quantify heterogeneity in a population of firms and derive
    a scaling relationship between firm size and the
    standard deviation of growth rates.
    Section~\ref{Tent} studies the distribution of
    annual growth rates for a single firm and
    approximates the growth distribution generated by a heterogeneous population.
    Section~\ref{OtherScaling} uses the scaling analysis
    presented in Sec.~\ref{FirmHeterogeneity}
    to illustrate the relationship between
    the number of products sold by firms and firm sales.
    \section{Model For The Statistics of a Single Firm}
    \label{StatisticsOfAFirm}
    We define the size of firm $\i$ to be its total annual
    sales $S_\i$ which is the sum of $N_\i$ transactions,
    \begin{equation}
    \label{SalesSum}
    S_\i \equiv T_{\i,1} + T_{\i,2} + \cdots + T_{\i,N_\i},
    \end{equation}
    where $S_i \geq 0$ and $T_{\i,k} > 0$ are measured in units of currency.
    We assume both $N_\i$ and $T_{\i,k}$ are
    independent random variables with finite moments.
    The total sales $S_\i$ is then a random variable.
%
%
%
    Firm $\i$ may increase its total sales by increasing its sales volume $N_\i$,
    or by selling more expensive products (increasing $T_\i$).
    The uncertainties in price and
    number depend on the nature of the products that a firm sells
    and the market.  We consider Poisson-distributed
    demand~\cite{Wu,Barbour,Dominey,NegativeBinomial},
    where $N_\i$ is described by a Poisson distribution, and we define
    \begin{equation}
    Q_i \equiv \Ev{N_\i} = \Var{N_\i}.
    \end{equation}
    With this choice, the size of a firm $S_\i$ is a random variable
    with a compound Poisson distribution, the details
    of which depend on the statistics of the individual
    transactions $T_\i$.

    For a single firm, the time-averaged mean and variance of sales
    are denoted by
    \begin{eqnarray}
    \label{RewriteMoments}
    \Ev{S_\i}   &\equiv& Q_\i \; X_\i   \nonumber \\
    \Var{S_\i}  &\equiv& Q_\i \; X_\i^2 c_\i.
    \end{eqnarray}
    where ${X_\i \equiv \Ev{T_\i}}$ is the mean transaction size and
    ${c_\i \equiv 1 + \Var{T_\i}/X_\i^2}$ is a measure of the statistical
    dispersion in transaction sizes.
    Because of its importance in the subsequent analysis,
    we introduce the logarithm of the firm size,
    \begin{equation}
    \label{LogSize}
    s_\i \equiv \log S_\i .
    \end{equation}
    The log size $s_\i$ is a random variable that is a function of the random variable $S_\i$.
    To simplify the notation, we also introduce the logarithms of the
    expected number of transactions
    \begin{equation}
    q_\i \equiv \log Q_\i,
    \end{equation}
    and the mean transaction size
    \begin{equation}
    x_\i \equiv \log X_\i.
    \end{equation}

    We estimate the first moment of $s_\i$ by computing the expected value
    of the Taylor series expansion of \myeq{LogSize} about the mean $\Ev{S_\i}$,
    \begin{equation}
    \Ev{s_\i} = \log\left( Q_\i \; X_\i \right) - \frac{c_\i}{2 Q_\i} + \cdots.
    \end{equation}
    In the limit that the number of transactions is large, ${Q_\i \gg 1}$, we retain only the leading order term,
    \begin{equation}
    \label{FirmMean}
    \Ev{s_\i} \approx \log\left( Q_\i \; X_\i \right) = q_i + x_i.
    \end{equation}
    The variance of $s_\i$ is estimated similarly,
    \begin{equation}
    \label{FirmVariance}
    \Var{s_\i} \approx  \frac{Q_\i \; \Ev{X_\i}^2 c_\i}{Q_\i^2 \Ev{X_\i}^2}
        = \frac{ c_\i}{Q_\i }.
    \end{equation}
    Importantly, the variance in the log size of a firm is inversely
    proportional to the number of transactions.
    We combine Eqs.~\eqref{LogSize},~\eqref{FirmMean}, and~\eqref{FirmVariance},
    to compactly express the (approximate) log size of a firm in terms of its mean
    and standard deviation,
    \begin{equation}
    \label{FirmLangevin}
    \begin{split}
    s_\i &= \Ev{s_\i} + \left ( \Var{s_\i} \,\right )^{1/2} \hat \eta \\
         &= q_\i + x_\i + \left(\frac{c_\i}{Q_i}\right)^{1/2}  \hat \eta,
    \end{split}
    \end{equation}
    where $\hat \eta$ is a random variable that has zero mean and unit variance.
    Equation~\eqref{FirmLangevin} expresses the log sales of a particular
    firm described by the three parameters $q_\i$, $x_\i$, and $c_\i$.

    To compute the statistics of sales growth,
    recall our assumption that the annual sales of firm $\i$ is
    statistically independent of its sales in the previous year.
    Let $s_\i$ and $s_\i'$ denote the
    log size of a firm in two consecutive years,
    then the annual growth is
    \begin{equation}
    \label{GrowthDefintion}
    g_\i = s_\i' - s_\i = \left(\frac{c_\i}{Q_i}\right)^{1/2} \left (\hat \eta' - \hat \eta \right).
    \end{equation}
    The growth $g_\i$ is a random variable with vanishing odd-integer moments
    and variance given by
    \begin{equation}
    \label{GrowthVariance}
    \Var{g_\i} = \frac{2 c_\i}{Q_i} = 2 c_\i \exp(-q_\i).
    \end{equation}

    Thus, we have defined a firm's total sales to be a
    sum of a random number of transactions of random size.  Each firm is uniquely described
    by its average sales volume and its transaction sizes,
    characterized by $q_\i$, $x_\i$, and $c_\i$.  The statistical independence of sales
    from year to year implies that the variance in growth rates,
    \myeq{GrowthVariance}, is a simple function of average sales volume
    and the transaction size dispersion.

    \section{Quantifying Heterogeneity in the Population of Firms}
    \label{FirmHeterogeneity}
    Prior studies~\cite{Nature96, Scaling1, Plerou2, BottazziPammolli, Fabritiis, Gaffeo}
    examined the statistics of sales
    growth within a population of firms and found
    that the standard deviation of growth rates scales as a
    power law with firm sales, $\sigma_g \sim \Size^{-\beta}$, and that
    the distribution of growth rates is non-Gaussian with slowly
    decaying tails.
    Analogously, we examine the statistics of sales growth for a
    population of firms.  We will find that
    the heterogeneity of firms is fundamental to reproducing
    the power-law scaling in the standard deviation of growth rates.

    We assume that an individual firm's parameters,
    $q_\i$, $x_\i$, and $c_\i$, evolve slowly and may be treated as
    fixed quantities.  To introduce heterogeneity within the population,
    these parameters are sampled from a time-independent
    distribution~\cite{Stationary}.
    In the subsequent discussion, we drop
    the subscript $\i$ if a
    parameter is a random variable that is representative of all
    firms (not one firm in particular.)
    Our analysis assumes that the parameter $c$
    has finite mean $\Ev{c}$, and that it is statistically independent of
    the other firm parameters.
    The means and variances of the parameters $q$ and $x$ across the population
    are denoted by
    \begin{align}
    \label{VolumeMoments}
    \Ev{q}   &= m_q   & \Var{q}   &= \sigma^2_q     \\
    \label{TransactionMoments}
    \Ev{x}      &= m_x & \Var{x} &= \sigma^2_x.
    \end{align}
%
    We take both $q$ and $x$ to be normally distributed
    within the population, with correlation coefficient $\rho$.
    The joint probability density function (PDF) of the pair is given by
    \begin{equation}
    \label{Bivariate}
    p(q, x) = \frac{\exp \left [ -\frac{1}{2} z (1-\rho^2)^{-1} \right ]}
                    {\sqrt{4 \pi^2 \sigma^2_q \sigma^2_x (1 - \rho^2)}},
    \end{equation}
    where
    \begin{equation}
    \begin{split}
    z \equiv    &\frac{(q - m_q)^2}{\sigma^2_q} + \frac{(x - m_x)^2}{\sigma^2_x} \\
                &\qquad - \frac{2 \; \rho \; (q - m_q)\;(x - m_x) }{\sigma_q \sigma_x}.
    \end{split}
    \end{equation}
    The correlation coefficient depends on the industry:
    a negative value of $\rho$ implies that firms that sell
    smaller quantities tend to sell more expensive items.
    The parameters of \myeq{Bivariate} also reflect the fact that firms
    must minimize inefficiency and risk.  Firms with too many transactions
    have excess overhead, and firms with low volume risk bankruptcy
    from large sales fluctuations.

    \begin{figure}
    \includegraphics[height=2.5in,width=3in]{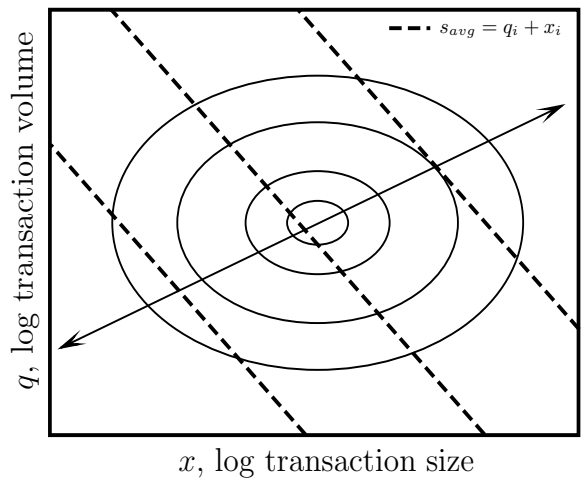}
    \caption{ \label{SchematicFigure}
    Solid ellipses are the level curves of the bivariate distribution
    given in \myeq{Bivariate} with $\rho = 0$.  Dashed lines show
    possible configurations of transaction volume and size for firms of various sizes.  The
    arrow indicates that the mean log transaction quantity increases
    linearly with log size, in agreement with \myeq{ConditionalMean}. }
    \end{figure}

    We examine the statistics
    of firms of a fixed mean size, $\avg{s} = q + x = const$.
    As illustrated in \myfig{SchematicFigure}, the relationship between
    total firm sales and transactions imply that statistics that depend on
    transaction volume $q_\i$ also depend on the sizes of the
    firms under consideration.
    For example, to extend the variance in growth rates given
    in \myeq{FirmVariance} to include all firms of a fixed size, we
    must determine the distribution of $q$ conditioned on $\avg{s}$,
    \begin{gather}
    \label{Bayes}
    p(q \given \avg{s} ) = \frac{p(q, \avg{s} )}{\int dq \, p(q, \avg{s} )},
    \end{gather}
    where $p(q, \avg{s} )$ is a Gaussian PDF that is obtained by
    substituting the constraint, ${x = \avg{s} - q}$, into \myeq{Bivariate},
    \begin{equation}
    p(q, \avg{s} ) = p(q, x = \avg{s} - q).
    \end{equation}

    The quotient of two Gaussian distributions
    is itself Gaussian.  Here, the conditional distribution in \myeq{Bayes} is
    completely characterized by its conditional mean
%
%
    \begin{equation}
    \begin{split}
    \label{ConditionalMean}
    \Ev{q \given \avg{s} } \; &= \; 2\beta \avg{s} \\
                              &  + \frac{(m_q - m_x) \rho \sigma_q \sigma_x + m_q \sigma^2_x - m_x \sigma^2_q}{u} \;,
    \end{split}
    \end{equation}
    and conditional variance
    \begin{equation}
    \label{ConditionalVariance}
    \Var{q \given \avg{s} } = \frac{(1-\rho^2)\sigma^2_q \sigma^2_x}{u},
    \end{equation}
    where
    \begin{equation}
    \label{BetaDefinition}
    \beta  = \frac{\sigma^2_q+\rho\sigma_q\sigma_x}{2 u},  \; \; u = \sigma^2_q + \sigma^2_x + 2\rho\sigma_q\sigma_x.
    \end{equation}

    We use Eqs.~\eqref{GrowthVariance},~\eqref{ConditionalMean}, and~\eqref{ConditionalVariance} to find the
    variance in annual growth rates averaged over all firms of a given log size,
    \begin{equation}
    \label{ConditionalGrowthVariance}
    \Var{g \given \avg{s} } = 2 \Ev{c} \exp \left [-\Ev{q \given \avg{s} } + \frac{\Var{q \, | \, \avg{s} }}{2}\right].
    \end{equation}
%
    This result recovers the empirically
    observed power-law relationship between firm sales $\avg{S} \equiv \exp(\avg{s})$
    and variability in growth rates,
    \begin{equation}
    \label{SigmaScaling}
    \Var{g \given \avg{S} } \sim \avg{S}^{\,-2 \beta} \quad \implies \quad \sigma_g \sim \avg{S}^{\,-\beta}.
    \end{equation}
%
    Among all firms of a fixed mean size, we find that
    the mean number of transactions per year scales as a power law,
    \begin{equation}
    \label{NumberScaling}
    \Ev{Q \given \avg{S} } \sim \avg{S}^{\,2\beta},
    \end{equation}
    where $Q \equiv \e^{q}$.  A similar calculation for the distribution 
    of the mean transaction
    size conditioned on firm size also gives a power law,
    \begin{equation}
    \label{SizeScaling}
    \Ev{X \given \avg{S} } \sim \avg{S}^{1-2 \beta},
    \end{equation}
    where $X \equiv \e^{x}$.

    Finally, we note that empirical data are frequently sparse and are
    binned into large size classes. Appendix~\ref{Binning} reviews how
    binning affects the conditional statistics
    presented in this section.
    \begin{figure*}
    \begin{minipage}[t]{3.2in}
    \includegraphics[height=2.5in,width=3in]{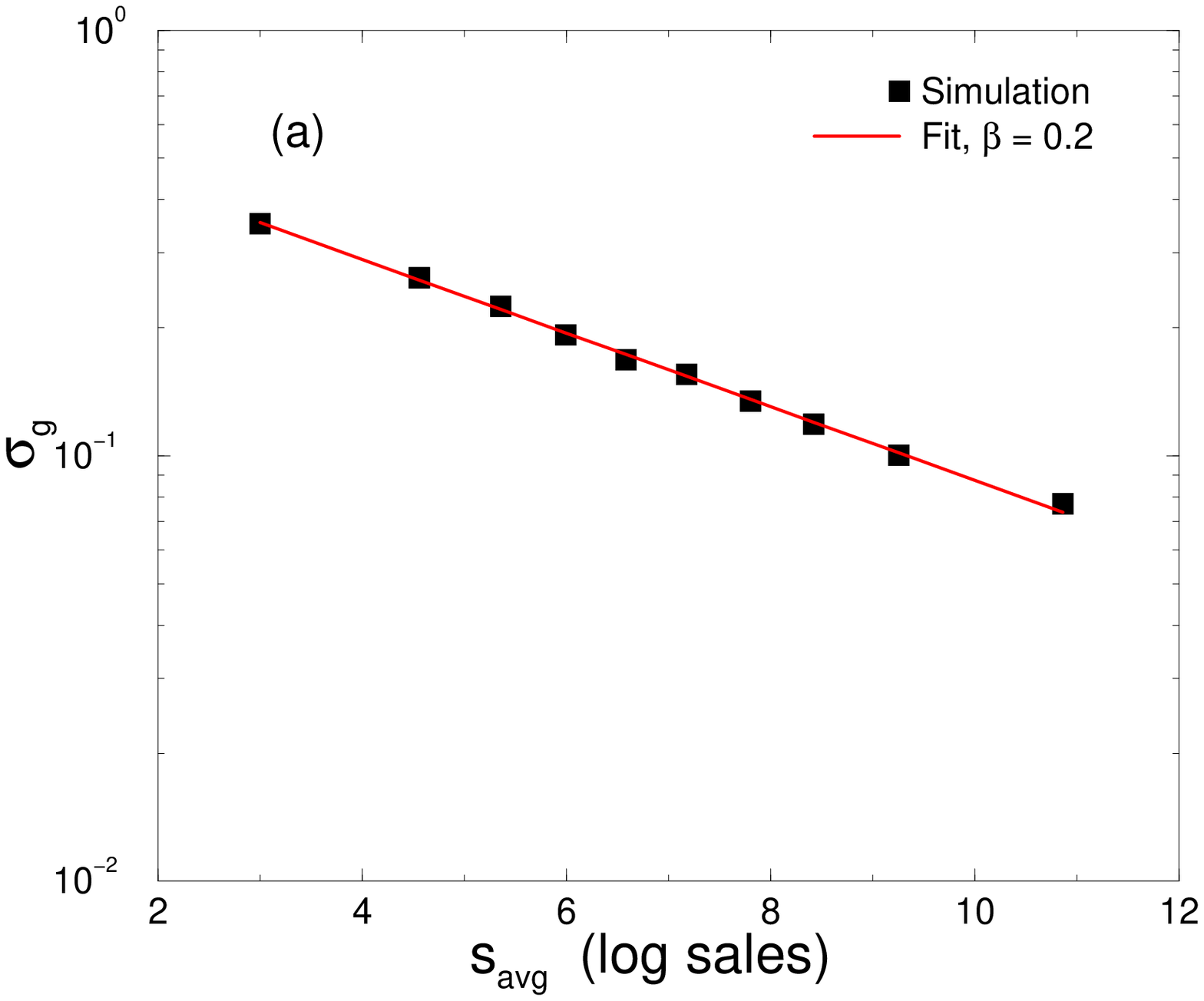}
    \end{minipage}
    \begin{minipage}[t]{0.1in}
    \hfill
    \end{minipage}
    \begin{minipage}[t]{3.2in}
    \includegraphics[height=2.5in,width=3in]{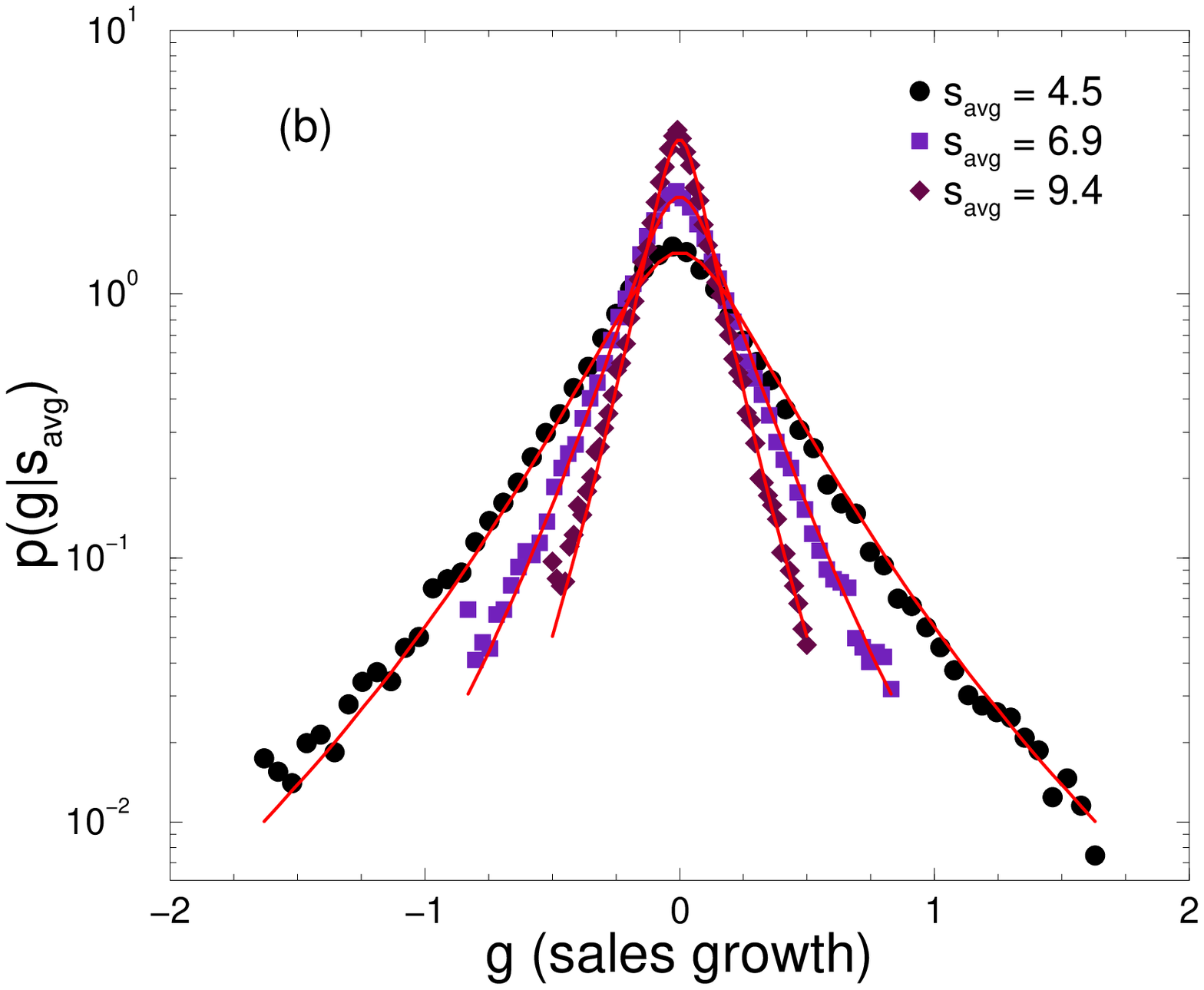}
    \end{minipage}
    \caption{ \label{SimulationFigure}
    Simulation of 3000 firms for 60 periods with $m_q   = 4.0$, $\sigma^2_q   = 2.0$,
    $m_x = 3.0$, $\sigma^2_x = 3.0$, $\rho = 0$, and constant $c = 1.01$.
    (a)  Power-law scaling of the standard deviation of growth rates
            conditioned on size, and fit to \myeq{SigmaScaling}.
    (b)  PDFs of growth rates for firms of
            three sizes and the corresponding approximate PDFs from \myeq{LaplaceGrowthPDF}. }
    \end{figure*}
\section{Distribution of Growth Rates}
\label{Tent}

    We next address the empirically observed non-Gaussian distribution
    of sales growth
    rates~\cite{Nature96,Scaling1,Plerou2,BottazziPammolli,Fabritiis,Gaffeo}.
    In general, a firm's sales growth distribution $p_{gi}$ depends on
    a set of firm-specific parameters, which we denote with the vector $\vec a_i$.
    For a set of heterogeneous firms $\cal F$,
    let $p_{a}(\vec a)$ denote the joint PDF of the parameters.
    Schematically, the growth rate distribution computed for all firms
    in the set $\cal F$ is given by
    \begin{equation}
    \label{Mixture}
    \begin{split}
    \Pi(g | {\cal F}) &\equiv \int d\vec a\; p_{gi}(g | \vec a) p_{a}(\vec a | {\cal F}) \\
                      &  =    \int_0^{\infty} d\sigma^2 p(g | \sigma^2) p(\sigma^2 | {\cal F}),
    \end{split}
    \end{equation}
    where the second equality is obtained by integrating over all degrees of freedom
    except the variance in growth rates, $\sigma^2$.
    For a sufficiently broad variance distribution $p(\sigma^2|{\cal F})$,
    the distribution of growth rates $\Pi(g|{\cal F})$
    is non-Gaussian with slowly decaying tails, a typical
    feature of growth models that incorporate heterogeneity
    of firms~\cite{PRL98,BottazziPhysica,Wyart,Fu,BottazziSecchi,BottazziSecchi2,Hanssen}.

    In the present treatment, we note that within a population of firms
    of fixed average annual sales, there are firms with large numbers of
    small transactions and {\it vice versa}.
    Because the variance in growth rates is inversely proportional
    to the number of transactions, the heterogeneity in transaction volume
    gives rise to a distribution in the variance of firm growth rates.
    To study the distribution of growth rates beyond the second moment,
    we must estimate the distribution of a firm's total sales $S_i$.

    In actuarial science and operations research literature~\cite{Dominey,MART},
    the distribution of $S_i$ has been approximated by the truncated normal,
    lognormal, and gamma distributions.
    For analytical convenience, we approximate the compound Poisson
    distribution of firm sales $S_i$ with a gamma distribution.
    As required by our theory,
    the gamma distribution admits only nonnegative sales, $S_i \geq 0$, and
    it is asymptotically Gaussian when the
    number of transactions is large, in agreement with the central limit theorem.
    The PDF of firm size $S_i$ is assumed to be
    \begin{equation}
    \label{GammaSize}
    p_{S\i} = \frac{S_\i^{a-1}}{\;b_\i^{a_\i} \; \Gamma(a_\i)} \, \exp\left(-\frac{S_\i}{b_\i}\right),
    \end{equation}
    where the parameters $a_\i$ and $b_\i$ are unique to firm $\i$.
    These parameters are determined by equating the mean and variance
    to their values given in \myeq{RewriteMoments},
    \begin{equation}
    \begin{split}
    \label{LinearEqs}
    \Ev{S_\i} &= Q_\i X_\i = b_\i a_\i, \\
    \Var{S_\i} &= Q_\i X_\i^2 c_\i = b_\i^2 a_\i.
    \end{split}
    \end{equation}
    We solve \myeq{LinearEqs} to obtain an expression
    for $a_\i$ in terms of the firm parameters,
    \begin{equation}
    a_\i = \frac{Q_\i}{c_\i} = \frac{\exp(q_\i)}{c_\i}.
    \end{equation}
    The PDF of a firm's log size $s_i$ follows from \myeq{GammaSize},
    \begin{equation}
    p_{s\i} = \frac{1}{\;b_\i^{a_\i}\; \Gamma(a_\i) } \exp[a_\i s_\i -\exp(s_\i)/b_\i].
    \end{equation}
    Equation~\eqref{GrowthDefintion} defines
    the growth of a single firm to be the difference of two
    independent random variables. Therefore, the PDF
    of single-firm growth rates is given by
    \begin{equation}
    \label{FirmGrowthRates}
    \begin{split}
    p_{g\i} & = \int_{-\infty}^{\infty} dg'\;p_{s\i}(g')\;p_{s\i}(g'+g_\i), \\
            & = \frac{\Gamma(2 a_\i + 1)}{a 2^{a+1} \Gamma^2(a_\i)} \left ( 1 + \cosh g \right )^{-a_\i} \\
            & = \frac{\Gamma(a_\i+\frac{1}{2})}{2 \sqrt{\pi} \; \Gamma(a_\i)} \cosh^{-2 a_\i} \left(\frac{g_\i}{2}\right).
    \end{split}
    \end{equation}
    In the case considered here, the distribution of single firm growth rates depends
    on a single scalar parameter, $a_i$.
    The variance in a firm's growth rate is given by
    \begin{equation}
    \begin{split}
    \Var{g_\i} =& \Ev{g_\i^2} = 2 \Psi_1(a_\i) \sim \frac{2}{a_\i} + \frac{1}{a_\i^2} + \cdots, \\
    \end{split}
    \end{equation}
    where $\Psi_1$ is the first derivative of the digamma function~\cite{AbramowitzStegun},
    \begin{equation}
    \Psi_1(x) = \frac{d}{dx} \Psi (x) = \frac{d^2}{dx^2} \log \Gamma(x).
    \end{equation}
    In the large-$a_i$ limit, we recover the expression
    for the variance in \myeq{GrowthVariance}.  This asymptotic agreement
    is expected because in Sec.~\ref{StatisticsOfAFirm} we employed
    a low-order expansion of the logarithm.
    Furthermore, because the hyperbolic cosine is asymptotically exponential, the tails
    of the distribution in \myeq{FirmGrowthRates} decay
    exponentially,

    The distribution of growth rates for a population
    must reflect the variability of the parameter $a$ among the firms.
    Formally, the PDF of growth rates for a population is given by
    a weighted mixture of the PDFs of single-firm growth rates given in \myeq{FirmGrowthRates}.
    To express this distribution analytically, we assume negligible
    variance in the transaction size dispersion, i.e., $c_\i = c$ is constant.
    In this case, the distribution of the parameter $a$ is lognormal,
    \begin{equation} 
    \label{ADensity}
    p_{a} = \frac{1}{a \sqrt{2 \pi V} } \exp\left[-\frac{1}{2V} \bigl (\log a \, - \, M \bigr)^2 \right],
    \end{equation}
    where $M \equiv \Ev{q | {\cal F} } - \log c$ and $V \equiv \Var{q | {\cal F}}$
    are determined by the statistics of ${\cal F}$, the set of firms under consideration.
    For example, to determine the growth rate distribution for the entire population,
    we use the moments defined in \myeq{VolumeMoments}.  Alternatively, to compute the distribution
    of growth rates among firms of fixed size, we use the conditional moments given
    in Eqs.~\eqref{ConditionalMean} and~\eqref{ConditionalVariance}.
    In the limit of many transactions, $a \gg 1$, we
    approximate the PDF of growth rates for a set of heterogeneous firms,
    \begin{equation}
    \label{LaplaceGrowthPDF}
    \Pi(g | {\cal F}) \approx \frac{\exp\left[ \bigl (4 V M + V^2 - 4 w^2 - 8 w\bigr) / \,8V \right ]}
        {\sqrt{4\pi(1+w)}},
    \end{equation}
    where
    \begin{equation}
    w = {\cal W}\left [2 V \log \left(\cosh \frac{g}{2}\right) \e^{M + V/2} \right],
    \end{equation}
    and ${\cal W}$ is the Lambert W function~\cite{Corless},
    defined by ${\cal W}(x) \e^{{\cal W}(x)} = x$.
    We give the details of the derivation and an asymptotic
    analysis of \myeq{LaplaceGrowthPDF} in Appendix~\ref{PDFDerivation}.

    Figure~\ref{SimulationFigure} compares
    \myeq{SigmaScaling} and \myeq{LaplaceGrowthPDF} to the
    statistics generated from a simulation of individual firms
    with annual sales sampled according to \myeq{GammaSize}.
    We see that our analysis produces a tent-shaped distribution, consistent
    with empirical facts~\cite{Nature96,Scaling1,Fabritiis}.
%
%
%
%
%

    \section{The Scaling of Products}
    \label{OtherScaling}

    Power-law scaling relationships are observed in
    academic, ecological, technological, and economic
    systems that relate a measure of size to the number
    of constituents~\cite{Plerou1,Preston,MacArthur,Keitt,Buldyrev,Goldenfeld,Lakhina}.
    For example, the number of different products sold by firms (the constituents)
    grows as a power law in firm sales (a measure of size), ${\Number \sim \Size^{0.42}}$~\cite{Matia2}.

    To elucidate the relationship between products and sales,
    we adopt the notation of Section~\ref{FirmHeterogeneity}.
    Let $Q_{\i}$ denote the number of products within firm $\i$
    and let $X_{\i}$ denote the firm's total annual sales per product.
    Because economic conditions are different in different locales and
    industries,
    we assume that among all firms, the number of products $Q$ and
    the mean sales per product $X$ are both lognormally
    distributed random variables.
    For convenience, we drop the subscript on the random variables
    and take the logarithm,
    \begin{equation}
    \label{ConditionalRelationship}
    S = Q X  \qquad\rightarrow\qquad s = q + x.
    \end{equation}
    The expression for the joint density of $q$ and $x$ is
    identical to \myeq{Bivariate}, except that in this context, the parameters
    of the distribution reflect aspects of the product portfolios of firms.

    From Eqs.~\eqref{NumberScaling} and~\eqref{SizeScaling},
    we find that the mean number of products and the mean product size scale
    as power laws with average firm size,
    \begin{eqnarray}
    \Ev{Q \given \avg{S} } &\sim& \avg{S}^{\,2\beta}, \\
    \Ev{X \given \avg{S} } &\sim& \avg{S}^{1 - 2\beta},
    \end{eqnarray}
    where $\beta$ is given in \myeq{BetaDefinition}.
    With a suitable choice of the parameters $\sigma^2_q$, $\sigma^2_x$
    and $\rho$, one can reproduce the observed power-law scaling
    relationship between the number of products and firm sales.
    Moreover, for $\rho = 0$ and with $\sigma_q^2 = {\cal W}$,
    and $\sigma_x^2 = {\cal D}$, the analysis here is identical to that
    of Refs.~\cite{PRL98,Matia2}.

%
%
%
%
%
%
%
%
%
%
%
%
    \section{Discussion}
%
%
%
%

    The present model of firm size fluctuations produces a
    non-Gaussian distribution of growth rates and generates a stationary
    lognormal distribution of firm sizes, features consistent with empirical studies.
    Our approach bears similarity to the model proposed 
    in Ref.~\cite{PRL98} in which a firm is
    split into subunits that each obey simple dynamics.  In the present work,
    the number of transactions is analogous to the number of subunits,
    and the transaction size parallels the sizes of the subunits within a firm.
%
%
%
    The model of Ref.~\cite{PRL98} postulates a complex
    internal structure within a firm.  In contrast,
    our approach decomposes a firm's sales into the individual transactions that
    occur in a year.

    Because transactions are well-defined measurable quantities,
    the assumptions and predictions of the theory are verifiable.
    In practice, correlations between transactions
    invalidate our assumption that the number of transactions $N_i$
    follows a Poisson distribution.
    Indeed, transactions within a firm can be regrouped to remove
    correlations.  Because these uncorrelated groups satisfy the assumptions
    of our model, we conclude that a Poisson-distributed
    number of transactions is not strictly necessary to obtain
    the results presented here.
%
%
    To retain our transactional framework, one can extend the model of a single firm's statistics
    described in Section~\ref{StatisticsOfAFirm}
    to account for correlations between transactions
    at the expense of additional parameters~\cite{NegativeBinomial}.
    Independent of the statistical details,
    most measures of microeconomic and macroeconomic consumption
    tally individual transactions and are therefore amenable to the
    transactional approach presented here.
    Figuratively speaking, transactions are the `atoms' of economic
    activity: they are discrete and are the basis
    of economies of all scales, from the individual to the national.

%
%
%
    Using our approach, we studied
    the relationship between the number of different
    products sold by firms and total sales receipts.  We
    found that differences between firms could account for the
    reported scaling relationship between the two quantities,
    $\Ev{Q \given \avg{S} } \sim \avg{S}^{\,2\beta}$.
    Our analysis applies to many systems
    where two distinct measures of size are related by
    a proportionality constant that varies because of economic,
    environmental, and other accidental circumstances.
    When the relationship between the measures is weak,
    the population is heterogeneous and non-trivial power
    laws emerge~\cite{Buldyrev}.
    For example, both transaction volume and total sales
    quantify the size of a firm, but no strict universal proportionality
    between these quantities exists.  In this case, \myeq{NumberScaling}
    predicts a power-law relationship between the quantities.
    In ecology, both island surface
    area and number of species quantify island size: islands with
    greater area generally harbor more species.  However,
    the mix of species and the number distinct niches varies
    from island to island.  In agreement with our hypothesis,
    empirical studies show that the number of species scales
    as a power law with island area
    $S \sim A^{z}$, $z \approx 0.2$~\cite{Preston,MacArthur,Goldenfeld}.
    Other studies have found that the density of Internet routers
    scale as a power law with population density, $R \sim P^{\alpha}$,
    $\alpha \approx 1.4$~\cite{Lakhina}.  In our treatment, the particular
    exponent is a function of the heterogeneity in the
    Internet architecture deployed within a study region.

    In summary, we have presented a theory of firm size fluctuations
    that explains a number of reported statistics in economics.
    The analysis also represents an important null hypothesis
    because it suggests that some emergent statistics are a consequence
    of heterogeneity in the population of firms.
    For these systems, understanding and quantifying heterogeneity
    seems to be the central problem in understanding macroscopic firm statistics.

    \section{Acknowledgments}
    We acknowledge discussions with K.~Matia, K.~Yamasaki, F.~Pammolli, and M.~Riccaboni.
    We also thank S.~Sreenivasan
    and George Schweiger for their reading of the manuscript
    and insightful suggestions. We thank DOE Grant No. DE-FG02-95ER14498
    and the NSF for support.

%
%
    \appendix
    \section{Statistics of Binning}
    \label{Binning}

    Frequently, empirical analysis of data cannot select
    a significant set of firms of a fixed
    size.  The usual remedy is to bin by selecting data within a range of sizes.
    Because the range may be large, the conditional
    statistics may change.  For example, we examine the
    conditional mean and variance of $q$,
    for the subset of firms in the bin $s_{\min} \leq \avg{s} \leq s_{\max}$.
    Within the bin, the distribution of $\avg{s}$ is a complicated
    function of $s_{\min}$ and $s_{\max}$.
    We denote the average of the log mean size of firms within the bin
    by $\Ev{\avg{s} \given B}$, the corresponding variance within
    the bin is denoted by $\Var{\avg{s} \given  B}$.
    The conditional expectation $\Ev{q \given \avg{s} }$ given in \myeq{ConditionalMean}
    is linear in $\avg{s}$,
    consequently, binning has no impact on the calculation of the conditional
    first moment.
%
%
    The variance of $q$ for the set of firms within the bin is given by
    \begin{equation}
    \Var{q \given B} = \Var{ q \given \Ev{\avg{s} \given B} }
        + 4 \beta^2 \Var{\avg{s} \given B},
    \end{equation}
    where $\Var{q \given \Ev{\avg{s} \given B} }$ is defined in \myeq{ConditionalVariance}.
    The above relationship implies that \myeq{SigmaScaling} holds only
    when we have binned firms such that the variance in firms' log mean size in each
    bin, $\Var{\avg{s} \given B}$, is a constant.
    Furthermore, because the mean and variance of $q$ is
    reflected in the parameters $M$ and $V$ in \myeq{ADensity},
    this implies that the distribution of growth rates also
    has a non-trivial dependence on bin width.

    \section{Estimation of the Growth Rate Distribution}
    \label{PDFDerivation}

    The formal expression for the PDF of growth rates
    for a set of firms $\cal F$ follows from Eqs.~\eqref{FirmGrowthRates} and~\eqref{ADensity},
    \begin{equation*}
    \label{PopGrowthDistribution}
    \Pi(g | {\cal F}) = \int da \frac{\e^{ -[\log a - M]^2/2V}}{a \sqrt{2 \pi V}}
        \frac{\Gamma(a+\frac{1}{2})}{2 \sqrt{\pi} \; \Gamma(a)} \cosh^{-2 a}\!\!\left(\frac{g}{2}\right).
    \end{equation*}
    In equilibrium, viable firms cannot sustain large sales
    fluctuations, so $a \gg 1$ and $M \gg V$.
    We replace the integrand with its leading-order term
    in the asymptotic expansion in $a$ and perform a change
    of variables,
    \begin{equation}
    \label{AprxPopGrowthDistribution}
    \begin{split}
    \Pi(g|{\cal F}) &\approx \int da \frac{\e^{ -[\log a - M]^2/2V}}{a \sqrt{2 \pi V}}
            \frac{\sqrt{a}}{2 \sqrt{\pi}}  \cosh^{-2 a}\!\!\left(\frac{g}{2}\right) \\
        & = \int dz \frac{\e^{z/2 -[z - M]^2/2V}}{2 \pi \sqrt{2 V} }
             \left [ \cosh \; \frac{g}{2} \right ]^{-2 \exp(z)}.
    \end{split}
    \end{equation}
    We approximate the second integral in \myeq{AprxPopGrowthDistribution}
    using the Laplace method to obtain \myeq{LaplaceGrowthPDF}.

    To examine the large-$g$ behavior of \myeq{LaplaceGrowthPDF},
    note that the Lambert W function is asymptotically logarithmic, ${\cal W}(x) \sim \log(x)$.
    We set $w \rightarrow \log g + \log V + M + V/2$ in \myeq{LaplaceGrowthPDF} to obtain,
    \begin{equation}
    \label{TailBehavior}
    \Pi(g | {\cal F}) \sim \frac{\exp[-\log^2\!g\,/\,2 V]}{\sqrt{\gamma_0 V  \, +  \, \log g\;}} \; g^{-\gamma_0}, \\
    \end{equation}
    where the exponent $\gamma_0$ is given by,
    \begin{equation}
    \gamma_0 = \frac{1}{2} + \frac{1 + M + \log V}{V}.
    \end{equation}
    For finite $V$, the tail of the distribution of growth rates $\Pi(g)$
    decays slower than an exponential, in agreement with empirical studies.


    \vfill


\begin{thebibliography}{98}
\bibitem{NegativeBinomial}
    One generalization is to let $N_\i$ be drawn from a
    negative binomial distribution.  In this case, we obtain a more
    general model at the expense of an additional parameter.
    More rigourous approaches are presented in Ref.~\cite{MART}.
\bibitem{Stationary}
    The assumption of a time-independent distribution is incompatible
    with changes in demand, technological innovation, and
    political and economic climate.  We assume that this
    assumption is a reasonable approximation
    for statistics gathered over tens of years.
\end{thebibliography}
\end{document}